\documentstyle[twoside,fleqn,espcrc2,epsfig]{article}

\def\slashchar#1{\setbox0=\hbox{$#1$}           
   \dimen0=\wd0                                 
   \setbox1=\hbox{/} \dimen1=\wd1               
   \ifdim\dimen0>\dimen1                        
      \rlap{\hbox to \dimen0{\hfil/\hfil}}      
      #1                                        
   \else                                        
      \rlap{\hbox to \dimen1{\hfil$#1$\hfil}}   
      /                                         
   \fi}                                         %

\def\bc{\begin{center}}
\def\ec{\end{center}}
\def\be{\begin{equation}}
\def\ee{\end{equation}}
\newcommand{\<}{\langle}
\renewcommand{\>}{\rangle}
\newcommand{\beq}{\begin{equation}}
\newcommand{\eeq}{\end{equation}}
\newcommand{\beqn}{\begin{eqnarray}}
\newcommand{\eeqn}{\end{eqnarray}}
\newcommand{\nn}{\nonumber}

\newcommand{\AmS}{{\protect\the\textfont2
  A\kern-.1667em\lower.5ex\hbox{M}\kern-.125emS}}

\title{The $U_A(1)$ Problem on the Lattice
\thanks{Talk presented by G.C. Rossi}}
\author {L. Giusti\address{Department of Physics, Boston University
        Boston, MA 02215 USA.},
G.C. Rossi\address{Theory Division, CERN, 1211 Geneva 23, 
	Switzerland}\thanks{On leave of absence from Dip. di Fisica, 
	Univ. di Roma ``{\it Tor Vergata}'', I-00133 Roma, Italy},
M. Testa\address{Dip. di Fisica, Univ. di  Roma ``{\it La Sapienza}'' 
	and INFN, Sezione di Roma 1, I-00185 Roma, Italy.},
G. Veneziano$^{\rm{b}}$}

\begin{document}

\begin{abstract}
\vspace{-.3cm}
If the expression of the topological charge density operator, suggested 
by fermions obeying the Ginsparg--Wilson relation, is employed, it is 
possible to prove on the lattice the validity of the Witten--Veneziano 
formula for the $\eta'$ mass.  Recent numerical results 
from simulations with overlap fermions in 2 (abelian Schwinger model) and 4 
(QCD) dimensions give values for the mass of the lightest 
pseudo-scalar flavour-singlet state that agree with theoretical 
expectations and/or experimental data.
\end{abstract}

\maketitle
\section{Introduction}
\label{sec:INTRO}
\vspace{-.1cm}
In a nut-shell the $U_A(1)$ problem lies in the fact that, in the 
absence of $U_A(1)$ anomaly contributions, the unphysical bound 
$m_{{\rm{octet}}, I=0} < \sqrt{3} m_{\pi}$ holds~\cite{WEIN}. 
In 1976 't Hooft~\cite{THOO} pointed out that the resolution of this
problem had to be related to the existence of topologically non-trivial gauge 
field configurations in Euclidean QCD. Dwelling on this
observation, it is possible to obtain an explicit formula for the $\eta'$ 
mass either in the 't~Hooft limit ($N_c\rightarrow\infty$, with $g^2 N_c$ 
and $N_f$ held fixed~\cite{NINF}), as was done in~\cite{WIT}, 
or by assuming that anomalous flavour-singlet axial 
WTIs retain their validity order by order in an expansion in 
$u\equiv N_f/N_c$ around $u=0$, as argued in~\cite{VEN}. In both cases 
one can derive the ``leading-order'' Witten--Veneziano (WV) relation
\beq
m_{\eta '}^2=\frac{2N_f}{F_\pi^2}A\, ,\quad
A=\int d^4x\, \<Q(x) Q(0) \>\Big{|}_{\rm{YM}}\, ,
\label{ETAMASS}\eeq
where $F_\pi\simeq 94\,{\rm{MeV}}$ is the pion decay constant, $A$ the 
``topological susceptibility" and $Q=\frac{g^2}{32\pi^2}
\epsilon_{\mu\nu\rho\sigma}{\rm{tr}}[F_{\mu\nu}F_{\rho\sigma}]$ the 
topological charge density. The notation $\<...\>|_{\rm{YM}}$
means that the $QQ$-correlator is to be computed in the pure Yang--Mills 
(YM) theory, i.e.~in the absence of quarks.

The formal relations in eq.~(\ref{ETAMASS}), when translated in any
regularized version of QCD, such as lattice QCD, become more involved 
and quite a number of subtleties have to
be dealt with in order to be able to determine the correct field theoretical 
expression of $A$, which should be used in the above equations.
Two problems need to be solved to make the formulae in~(\ref{ETAMASS})  
precise and of practical use. One has to i) find a properly normalized 
lattice definition of the topological charge density, $Q_L$;
ii) subtract from the $Q_L(x)Q_L(0)$ product appropriate
contact terms, as required to make it an integrable (operator-valued)
distribution.

A rigorous derivation of a formula for the $\eta'$ mass, which could be 
unambiguously employed in numerical simulations, can be given~\cite{GRTV1}, 
if use is made of the (lattice) regularized anomalous flavour-singlet 
axial WTIs of full QCD in order to  construct a properly defined $\<Q_LQ_L\>$ 
correlation function. The remarkable result 
of this analysis is that no subtraction is needed to this end in 
the chiral limit, if the definition of lattice topological charge 
density suggested by fermions obeying the Ginsparg--Wilson (GW) 
relation~\cite{GW} (e.g.~overlap fermions~\cite{NEUB}) is employed.
\vspace{-0.2cm}
\section{GW fermions and the $\eta'$ mass formula}
\label{sec:GWF}
\vspace{-0.2cm}
Regularizing the fermionic part of the QCD action using GW
fermions~\cite{GW,NEUB} offers the great advantage over the standard
Wilson discretization~\cite{W} that global chiral transformations can be 
defined, which are an exact symmetry of the massless theory, as in the formal  
continuum limit. This follows from the relation
$\gamma_5D+D\gamma_5=D\gamma_5D$~\cite{GW}, obeyed by the GW Dirac 
operator, $D$~\footnote{We set the lattice spacing, $a$, equal to 1. 
In all lattice formulae the continuum limit is, however, understood.}. 
In this regularization the $U_A(1)$ 
anomaly is recovered \`a la Fujikawa~\cite{F}, because of the 
non-trivial Jacobian that accompanies the change of fermionic integration 
variables induced by a $U_A(1)$ lattice transformation. As a consequence the 
(anomalous) flavour-singlet WTIs in the presence of $N_f$ massless fermions 
take the form
\begin{eqnarray*}
\nabla_\mu\<A^0_\mu (x) \hat{O}\>=\frac{2N_f}{2}
\<{\rm{Tr}}[\gamma_5 D(x,x)] \hat{O}\>+\<\delta^x_A \hat{O}\>\, ,
\end{eqnarray*}
where $\hat{O}$ is any finite (multi)local operator, $\delta^x_A \hat{O}$ 
is its local variation and $\nabla_\mu A^0_\mu (x)$ is the 
divergence of the singlet axial current.

Neither ${\rm{Tr}}[\gamma_5 D(x,x)]$ nor $A^0_\mu (x)$ are finite operators, 
but finite linear combinations, $\hat Q$ and $\hat A^0_\mu$, can be
easily constructed~\cite{BARD} by writing
\begin{eqnarray}
&&\hat{Q}(x)=\frac{1}{2}{\rm{Tr}}[\gamma_5 D(x,x)] - 
\frac{Z}{2N_f}\nabla_\mu A_\mu^0 (x)
\label{FINQ1}\\
&&\hat{A}_\mu^0 (x)=(1-Z) A_\mu^0 (x)\, .
\label{FINA} \end{eqnarray}
where $Z$ is logarithmically divergent to lowest order in perturbation 
theory and vanishes as $u\rightarrow 0$.

With the above definitions the renormalized singlet axial WTI becomes
\begin{equation}
\nabla_\mu\<\hat A^0_\mu (x) \hat{O}\>=
2N_f\<\hat Q(x) \hat{O}\>+\<\delta^x_A \hat{O}\> \, .
\label{WTIREN}
\end{equation}
This equation shows that $\hat Q$ and $\hat A^0_\mu$ are 
correctly normalized and that there is no power-divergent mixing of 
${\rm{Tr}}[\gamma_5 D(x,x)]$ with the pseudo-scalar quark density, which would
bring in a dangerous lineraly divergent mixing coefficient.

A formula for the $\eta'$ mass can be obtained from the above WTIs, 
by observing that the $\eta'$ mass must vanish as $u\rightarrow 0$ for 
massless QCD to have no $\theta$-dependence. 
For that one starts by defining the lattice Green function~\footnote{A 
contact term, ${\rm{CT}}(p)$, should be added to the r.h.s. of 
eq.~(\ref{AWTI2}) to make it finite at $p\neq0$. ${\rm{CT}}(p)$ is a 
polynomial of degree 4 in $p$, which vanishes at $p=0$, because the r.h.s. 
of eq.~(\ref{AWTI2}) is certainly finite (actually 
zero) at $p=0$. ${\rm{CT}}(p)$ plays no r\^ole in the argument below, since 
we will be finally only interested in the value of $\chi_{tL}$ at $p=0$. For 
brevity we will not indicate it explicitly in the following.}
\begin{equation}
\chi_{tL}(p)=\frac{1}{2N_f}\int d^4x\, {\rm{e}}^{-ipx} 
\nabla_\mu\<\hat A^0_\mu (x) \hat Q(0)\>\, ,
\label{AWTI2}
\end{equation}
with $\hat{Q}$ given by eq.~(\ref{FINQ1}). In the full theory, where the 
$\eta'$ is massive, one has $\chi_{tL}(0)=0$. On the other hand, since, as we
obseved above, at the chiral point, the $\eta'$ mass vanishes as 
$u\rightarrow 0$, only the $\eta'$ pole will contribute to $\chi_{tL}(p)$ 
in this limit, leading to the relation
\begin{equation}
\lim_{p\rightarrow 0}\lim_{u\rightarrow 0} \chi_{tL}(p)=
\frac{F_\pi^2}{2N_f} m_{\eta'}^2\Big{|}_{u=0}\, .
\label{AWTI4}
\end{equation}
The l.h.s. of eq.~(\ref{AWTI4}) can be evaluated by using 
the WTI~(\ref{WTIREN}) with $\hat{O}=\hat{Q}$. Recalling that $Z$ vanishes 
when $u\rightarrow 0$ and $\delta^x_A\hat{Q}=0$, we obtain
\begin{eqnarray}
&&\frac{F_\pi^2}{2N_f} m_{\eta'}^2\Big{|}_{u=0}=
\lim_{p\rightarrow 0}\lim_{u\rightarrow 0}
\int d^4x \,{\rm{e}}^{-ipx}\nn\\
&&\<\frac{1}{2}{\rm{Tr}}[\gamma_5 D(x,x)]
\frac{1}{2} {\rm{Tr}}[\gamma_5 D(0,0)]\>\, .
\label{ETAINTER}\end{eqnarray}
The limits $u\rightarrow 0$ and $p\rightarrow 0$ in this expression can be 
readily performed in the order indicated, if one can assume that taking 
the first limit simply amounts to setting the fermion determinant equal 
to unity. One gets in this way
\begin{eqnarray}
&\!\!\!\!\!\!\!&\frac{F_\pi^2}{2N_f} m_{\eta'}^2\Big{|}_{u=0}= 
\nn\\ &\!\!\!\!\!\!\!&=
\int d^4x \<\frac{1}{2}{\rm{Tr}}[\gamma_5 D(x,x)]
\frac{1}{2} {\rm{Tr}}[\gamma_5 D(0,0)]\>\Big{|}_{\rm{YM}}\, .
\label{ETAMASSOVER}\end{eqnarray}
The restriction to pure YM theory, indicated in eq.~(\ref{ETAMASSOVER}), 
is an obvious consequence of the fact that, for a Green function 
of only gluonic operators, neglecting the fermion determinant is 
tantamount to limiting the functional integral to the pure gauge sector 
of the theory. 

A relevant question at this point is to ask at which value of $N_c$ 
eq.~(\ref{ETAMASSOVER}) is supposed to be valid. The answer depends on the 
behaviour of QCD with $N_f$. Various scenarios are envisegeable. 

1) In the most favorable situation, in which the limit 
$u\rightarrow 0$ of $\chi_{tL}(p)$ exists and is equal to the value 
it takes at $N_f=0$ (fermion determinant equal to 1),
formula~(\ref{ETAMASSOVER}) is valid for any value of $N_c$ and for each 
$N_c$ it yields the mass of the $\eta'$ meson (at leading order in $u=0$) 
in the world with the corresponding number of colours~\cite{VEN}. 

2) If quenching can be attained only in the limit in which the number of 
colours goes to infinity, then eq.~(\ref{ETAMASSOVER}) will yield a 
formula for the $m_{\eta'}^2$ valid up to O($1/N_c$) corrections~\cite{WIT}.

3) Finally it may happen that taking the limit $u\rightarrow 0$ does not 
correspond to quenching. In this case one cannot pass from the fairly 
complicated eq.~(\ref{ETAINTER}) to the more useful 
formula~(\ref{ETAMASSOVER}). 

Remembering that $\frac{1}{2} {\rm{Tr}}[\gamma_5 D(x,x)]$ should be 
identified with the topological charge density~\cite{NEUB}, we conclude 
that eq.~(\ref{ETAMASSOVER}) can be rewritten in the very suggestive form
\beq
\frac{F_\pi^2}{2N_f} m_{\eta'}^2\Big{|}_{u=0}=
\lim_{V\rightarrow\infty}\frac{\<(n_R-n_L)^2\>}{V}\, ,
\label{ETAMASSIND}\eeq
where $\<(n_R-n_L)^2\>$ is the expectation value of the square of the 
index of the GW fermion operator, $D$, and $V$ is the physical volume 
of the lattice.

In the form~(\ref{ETAMASSIND}) the $\eta'$ mass formula can be directly
compared with overlap fermion simulation data. Existing numerical results 
in 2 (abelian Schwinger model~\cite{GHR}) and in 4 (QCD~\cite{EHN}) 
dimensions agree quite well with theoretical expectations and/or 
experimental numbers~\cite{GRTV1}.
\vspace{-0.2cm} 
\section{Conclusions}
\label{sec:CONCL}
\vspace{-0.2cm}
We have shown that a formula for the $\eta'$ mass can be rigorously derived 
in lattice QCD, exploiting the anomalous flavour-singlet axial WTIs of 
the theory. If fermions obeying the GW relation are used, there exists a 
natural definition of topological density which is correctly
normalized and for which the naive form of the WV formula holds without 
the need of introducing any subtraction. 

In the literature two sorts of approaches have been proposed to deal with 
the problem of computing $A$ on the lattice, which have led to numerical 
values as good as those obtained with the present method~\cite{GRTV1}. 
The first one is based on a direct 
field theoretical definition of $A$~\cite{DIG} that takes into account the 
need for the renormalization of $Q_L$ and the subtraction of the operators 
$F^2$ and $1\!\!\!\! 1$ in the short distance expansion of $Q_LQ_L$. The 
second one makes use of the notion of ``cooling''~\cite{DFGPS} to carry out 
the necessary operations of renormalization and subtraction. Both methods are 
reliable to the extent that they are able to capture the topological 
properties of the gauge field configurations that determine the number of 
zero modes of the Dirac operator. Simulations based on the geometrical 
definition of $Q_L$ of ref.~\cite{LU1} have not yet led to comparably good 
results~\cite{SCHI}.
\vskip .1cm
{\bf Acknowledgements} - I wish to thank the Organizers of Lattice2001 
for the wonderful and exciting atmosphere of the Conference.

\end{document}